\documentclass[aps,prd,notitlepage,superscriptaddress,longbibliography,nofootinbib]{revtex4-2}
\usepackage[dvipdfmx]{graphicx}

\usepackage{dcolumn}    
\usepackage{bm}         
\usepackage{hyperref}   
\usepackage{bmpsize}
\usepackage[english]{babel}
\usepackage[skip=1pt, justification=raggedright]{caption}
\usepackage{graphicx}
\vspace{-10pt}
\usepackage{amsmath}
\usepackage{subcaption}
\hypersetup{
    colorlinks=true,
    citecolor=blue,
    linkcolor=blue,
    urlcolor=blue
}

\begin{document}

\title{Noise budget of Cryogenic sub-Hz cROss torsion-bar detector with quantum NOn-demolition Speed meter (CHRONOS)
}

\author{Mario Juvenal S. Onglao III}
\thanks{Corresponding author: monglao@nip.upd.edu.ph}
\affiliation{National Institute of Physics, University of the Philippines - Diliman, Quezon City 1101, Philippines}
\affiliation{Center for High Energy and High Field (CHiP), National Central University,  Taoyuan, Taiwan}

\author{Hsiang-Yu Huang}
\affiliation{Department of Physics,National Central University, Taoyuan, Taiwan}
\affiliation{Center for High Energy and High Field (CHiP), National Central University,  Taoyuan, Taiwan}

\author{Yuki Inoue}
\affiliation{Department of Physics,National Central University, Taoyuan, Taiwan}
\affiliation{Center for High Energy and High Field (CHiP), National Central University,  Taoyuan, Taiwan}
\affiliation{Institute of Physics, Academia Sinica, Taipei, Taiwan}
\affiliation{Institute of Particle and Nuclear Studies, High Energy Acceleration Research Organization (KEK), Tsukuba, Japan}

\author{Vivek Kumar}
\affiliation{Department of Physics,National Central University, Taoyuan, Taiwan}
\affiliation{Center for High Energy and High Field (CHiP), National Central University,  Taoyuan, Taiwan}

\author{Daiki Tanabe}
\affiliation{Institute of Physics, Academia Sinica, Taipei, Taiwan}
\affiliation{Center for High Energy and High Field (CHiP), National Central University,  Taoyuan, Taiwan}
\affiliation{Institute of Particle and Nuclear Studies, High Energy Acceleration Research Organization (KEK), Tsukuba, Japan}

\begin{abstract}
CHRONOS is a proposed gravitational-wave detector designed to operate in the sub-Hz frequency (0.1–10 Hz) band, a largely unexplored range due to strong noise sources that hamper ground-based detectors. It employs cryogenic operation, cross torsion-bar configuration, triangular Sagnac interferometer, and speed meter readout scheme to overcome key noise limitations, targeting a strain sensitivity of $h \sim 10^{-18}\,\mathrm{Hz^{-1/2}}$ around $2\,\mathrm{Hz}$ and a stochastic gravitational wave background of $\Omega_{GW} \sim 2\times 10^{-3}$ at $2\,\mathrm{Hz}$. Using analytical and interferometric simulations with FINESSE3, we evaluate the noise budget of CHRONOS and characterize the relative contributions of quantum, thermal, and environmental noise sources. Our results demonstrate that CHRONOS achieves competitive sensitivity at low frequencies. The feasibility of using CHRONOS in an earthquake early-warning system by detecting prompt gravity-gradient signals is also investigated, and is predicted to be faster by approximately 2.92 to 6.90 seconds within 40 km. These findings highlight the scientific potential of CHRONOS, bridging gravitational-wave astronomy and geophysical monitoring, and motivating further development of low-frequency detector technologies.
\end{abstract}

\maketitle
\section{Introduction}\label{sec:intro}
Gravitational waves are ripples in spacetime produced by accelerating massive objects and were first directly detected in 2015 by Advanced LIGO \cite{abbott2016observation}. This has opened a new window into the universe, enabling the study of strong-field gravity theories, compact object populations, and cosmological processes that are otherwise inaccessible through electromagnetic observations.

Current ground-based detectors, notably Advanced LIGO, Virgo, and KAGRA, have now approached their design sensitivity \cite{buikema2020sensitivity,capote2025advanced}. This milestone has enabled routine detections and the establishment of gravitational-wave event catalogs. However, further improvements in sensitivity are necessary to probe weaker and more distant sources and to access other frequency bands. Motivated by these goals, several next-generation (third-generation) detectors are under development, including Cosmic Explorer \cite{reitze2019cosmic}, the Einstein Telescope \cite{punturo2010einstein}, as well as alternative designs such as torsion-bar antennas \cite{ando2010torsion,mcmanus2017mechanical} and other low-frequency instruments. These detectors aim to extend both the sensitivity and bandwidth of gravitational-wave observations, particularly toward the sub-Hertz regime.

Cryogenic sub-Hz cROss torsion-bar detector with quantum NOn-demolition Speed meter (CHRONOS) \cite{inoue2026opticaldesignsensitivityoptimization} is a proposed detector targeting the low-frequency range of 0.1–10 Hz at a design strain sensitivity of $h \sim 10^{-18}\,\mathrm{Hz^{-1/2}}$ around $2\,\mathrm{Hz}$. It incorporates key technologies, such as cryogenic operation to suppress thermal noise, cross torsion-bar configuration to enhance sensitivity to differential rotation, and triangular Sagnac interferometer with speed meter readout to reduce quantum back-action noise, to enable high-sensitivity measurements. For a complete discussion of the setup, the reader is referred to Ref. \cite{inoue2026opticaldesignsensitivityoptimization}. 

One such application of CHRONOS is in earthquake early detection. Rapid mass redistribution from fault rupture produces a gravitational disturbance that travels at the speed of light and can be detected before the slower seismic P-waves arrive \cite{harms2013low}. Such prompt gravitational signals, which have traditionally been treated as noise, can instead be exploited as a signal of interest.

In this work, we evaluate the sensitivity of the CHRONOS detector using analytical and full interferometric simulations performed with FINESSE3, and quantify its capability to detect prompt gravitational signals from earthquakes as part of an early warning system. This approach enables a direct assessment of CHRONOS not only as a gravitational-wave detector, but also as a candidate instrument for gravity-based early warning systems.

\section{Noise Budget and Sensitivity}\label{sec:noise}
Gravitational-wave detectors are limited by a combination of quantum (shot and radiation pressure), thermal (coating and bar), and environmental (seismic and Newtonian) noise, each dominant in different frequency regimes. For brevity, we summarize the key expressions derived in Ref. \cite{inoue2026chronosscienceprogram}. A complete derivation is provided in that work.

Shot noise arises from the discrete nature of photons and manifests as fluctuations in the number of output photons detected, dominating at high frequencies \cite{caves1981quantum}. In contrast, radiation pressure noise originates from fluctuations in the momentum transfer of photons to the end test masses (ETM), introducing uncertainty in the position \cite{caves1981quantum}, dominating at low frequencies. Utilizing a speedmeter readout scheme allows partial cancellation of radiation pressure noise at low frequencies. These effects were simulated using FINESSE3.

Coating Brownian noise arises from mechanical losses in the mirror coatings, which is particularly significant in the mid-frequency band \cite{harry2002thermal}. Eq.~\ref{eqn:coating} describes the amplitude spectral density of the displacement noise, where $\eta_g/L_{\text{bar}}$ is the geometric conversion factor from physical ETM displacement to interferometer readout, $k_B$ is Boltzmann's constant, $T$ is the coating temperature, $\sigma_s$ and $Y_s$ are the Poisson ratio and Young's modulus of the ETM substrate, respectively, $\omega_{\text{ETM}}$ is a geometric factor based on the beam spatial profile, and $\phi_c^{\text{eff}}$ is the effective mechanical loss angle of the coating stack. The noise scaling as $\sqrt{T}$ motivates cryogenic operation to mitigate coating Brownian noise.
\vspace{-0.3cm}
\begin{equation}
S_{\text{coat}}(\Omega)=\frac{\eta_g}{L_{\text{bar}}}\sqrt{\frac{4k_BT}{\pi \Omega}}\sqrt{\frac{1-\sigma_s^2}{\omega_{\text{ETM}}Y_s}}\sqrt{\phi_c^{\text{eff}}}
\label{eqn:coating}
\end{equation}
Bar thermal noise is associated with thermal fluctuations that excite the torsional modes of the bars, leading to angular displacement noise, especially near resonance frequencies. Eqs.~\ref{eqn:barbg}-\ref{eqn:barrot} describe the background and rotational bar thermal noise, where $\phi_s^{\text{eff}}$ is the effective material loss angle of the bar, $\omega_{tb}$ is the torsional resonance angular frequency, $Q=1/\phi_0$ is the mechanical quality factor, $I$ is the torsion bar moment of inertia, and other variables are the same as in Eq.~\ref{eqn:coating}. The combined bar thermal noise can be obtained via a quadratic sum: $S_{\text{bar}}(\Omega)=\frac{\eta_g}{L_{\text{bar}}}\sqrt{S_{\text{bar,bg}^2(\Omega)}+S_{\text{bar,rot}}^2(\Omega)}$. Reducing mechanical loss, careful selection of resonance frequencies, and operating at low temperatures are key strategies to mitigating this effect.
\vspace{-0.3cm}
\begin{eqnarray}
\label{eqn:barbg}
S_{\text{bar,bg}}(\Omega)&=&\sqrt{\frac{4k_BT}{\sqrt{\pi}\Omega}}\sqrt{\frac{1-\sigma_s^2}{Y_s}}\sqrt{\phi_s^{\text{eff}}} \\
\label{eqn:barrot}
S_{\text{bar,rot}}(\Omega)&=&\sqrt{\frac{4k_BT}{I}}\sqrt{\frac{\Omega \omega_{tb}/Q}{(\omega_{tb}^2-\Omega^2)^2+(\omega_{tb}\Omega/Q)^2}}
\end{eqnarray}
Seismic noise arises from ground vibrations due to natural anthropogenic activity and dominates at very low frequencies. Eq.~\ref{eqn:seismic} shows the strain sensitivity after attenuation by the suspension system. $S_x^{\text{ground}}$ is the measured ground-displacement spectrum at the proposed site, $H_{\text{yaw}}(\Omega)$ is the suspension yaw transfer function, $G_{\mu g}$ is the conversion gain between mechanical torque and gravity acceleration, and $G_{\text{pre}}$ is the attenuation of the pre-isolation system. To mitigate these effects, CHRONOS plans to employ active and passive vibration-isolation platforms and cryogenic suspensions. 
\vspace{-0.3cm}
\begin{equation}
S_{\text{seis}}(\Omega)=\frac{|H_{\text{yaw}}(\Omega)|}{G_{\mu g}G_{\text{pre}}}S_x^{\text{ground}}(\Omega)
\label{eqn:seismic}
\end{equation}

Although advanced isolation systems can attenuate seismic disturbances, changes in the environmental mass density can still gravitationally couple with the detector, leading to residual motion and gravity-gradient (Newtonian) noise that remain major challenges below a few hertz \cite{harms2013low}.

\begin{figure}[h!]
\centering
\begin{subfigure}{0.4\linewidth}
    \centering
    \includegraphics[width=\linewidth]{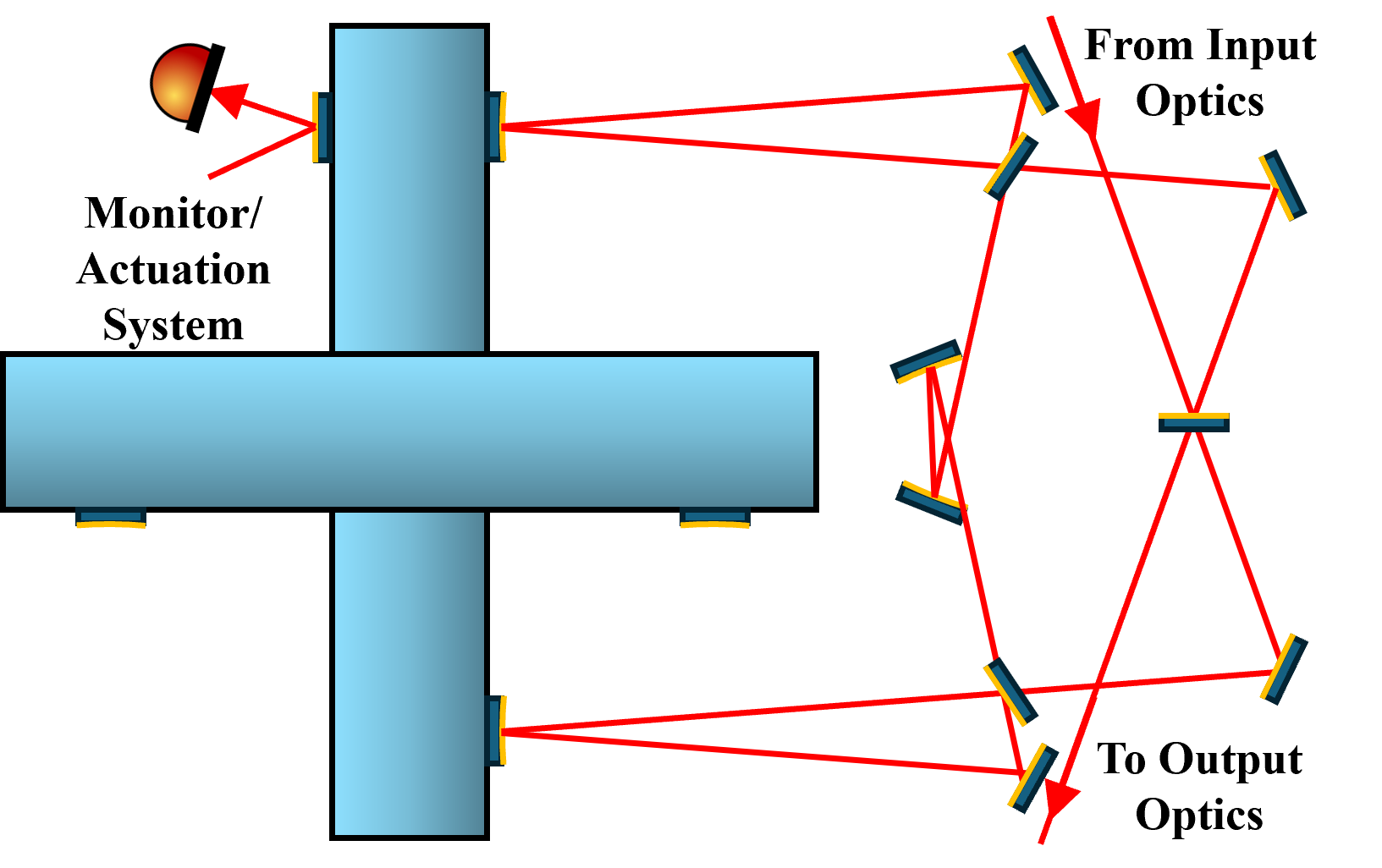}
    \caption{}
    \label{fig:CHRONOS}
\end{subfigure}
\hfill
\begin{subfigure}{0.4\linewidth}
    \centering
    \includegraphics[width=\linewidth]{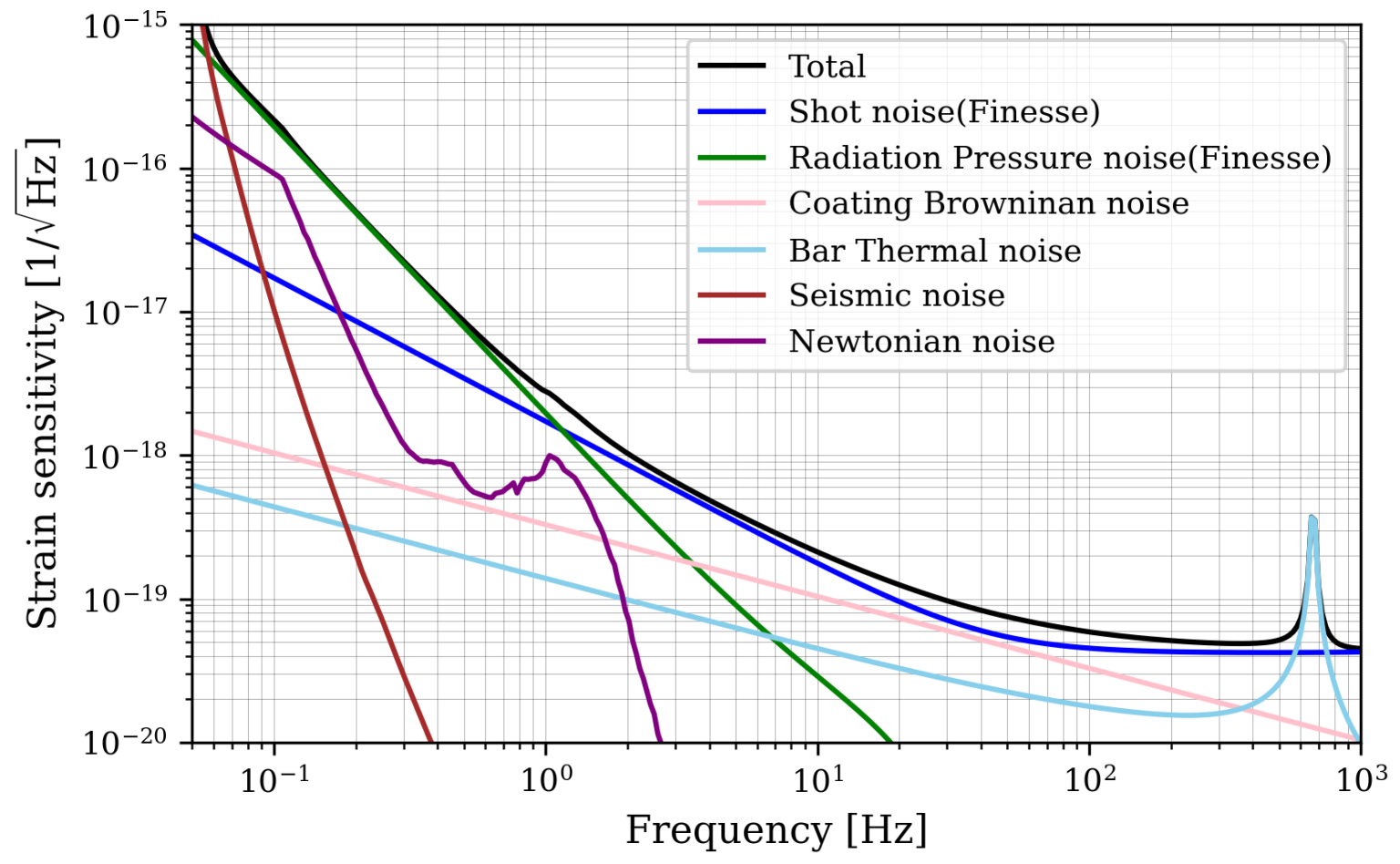}
    \caption{}
    \label{fig:sensitivity}
\end{subfigure}
\caption{(a) Simplified CHRONOS design. (b) Projected strain sensitivity of optimized CHRONOS(2.5 m). The black curve shows the overall sensitivity. Separate curves indicate the contributions from shot noise, radiation-pressure noise, coating Brownian noise, torsion-bar thermal noise, seismic noise, and Newtonian noise due to Rayleigh waves.}
\label{fig:EQ}
\end{figure}

Fig.~\ref{fig:CHRONOS} shows a simplified design of CHRONOS featuring the Sagnac interferometer and actuation system, while Fig.~\ref{fig:sensitivity} shows the optimized sensitivity of CHRONOS at a 2.5 m detector scale. CHRONOS sensitivity is limited by different noise sources across frequency bands: shot noise from optical phase fluctuations dominates at high frequencies, while radiation-pressure noise rises at lower frequencies but is mitigated by the Sagnac speed-meter configuration. In the 1–10 Hz range, coating Brownian noise sets the limit, marking the transition between quantum and thermal noise regimes. At sub-Hz frequencies, torsion-bar rotational thermal noise becomes dominant due to direct measurement of rotational motion, and at the lowest frequencies, Newtonian noise from Rayleigh-wave-induced fluctuations imposes the fundamental environmental limit, as it cannot be reduced by mechanical isolation. Understanding and mitigating these noise sources is essential for improving detector sensitivity and extending the observable frequency range of gravitational-wave instruments.


\section{Earthquake-induced prompt gravity signal}\label{sec:earthquake}
Earthquake-induced gravity gradients can be modeled as prompt gravitational perturbations arising from mass redistribution. Averaging over detector orientations, this is expressed as $\langle h_+(\bm{r_0},t) \rangle = \langle h_\times(\bm{r_0},t) \rangle =\frac{6\sqrt{14/5}~G}{r_0^5}I_4[M_0](t)$, where $h_{+}$ and $h_\times$ are the gravity strain projections based on the ETM distance changes and bar relative rotation, respectively, $G$ is the gravitational constant, $r_0$ is the distance to the earthquake centroid, and $I_4[M_0](t)$ is the fourth time integral of the seismic moment time function \cite{harms2015transient}.

Figure~\ref{fig:EQ} presents the predicted sensitivity of CHRONOS at 2.5 m detector scale (solid black curve), together with simulated prompt gravitational signals from earthquakes at various detector distances (colored dashed curves). The vertical axis corresponds to the amplitude spectral density (ASD) of the induced gravitational perturbations, while the horizontal axis shows frequency. Colored dashed lines represent the expected gravity-gradient signal for an earthquake with moment magnitude Mw $5.2$, based on models of rapid mass redistribution during fault rupture. A moment magnitude of 5.2 was selected as the minimum threshold of engineering significance in the Philippines \cite{thenhaus1994estimates}.

\begin{figure}[h!]
\centering
\begin{subfigure}{0.45\linewidth}
    \centering
    \includegraphics[width=\linewidth]{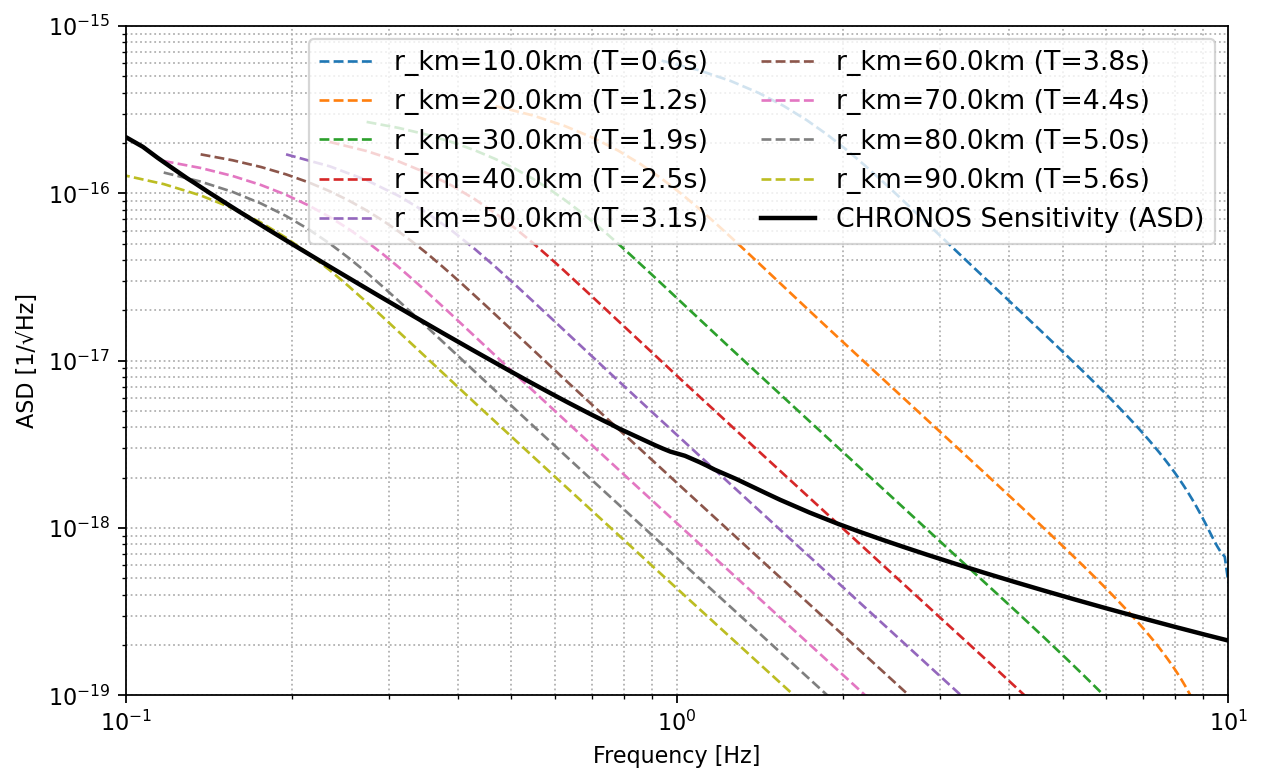}
    \caption{Sensitivity with prompt gravity signals.}
    \label{fig:EQ_a}
\end{subfigure}
\hfill
\begin{subfigure}{0.27\linewidth}
    \centering
    \includegraphics[width=\linewidth]{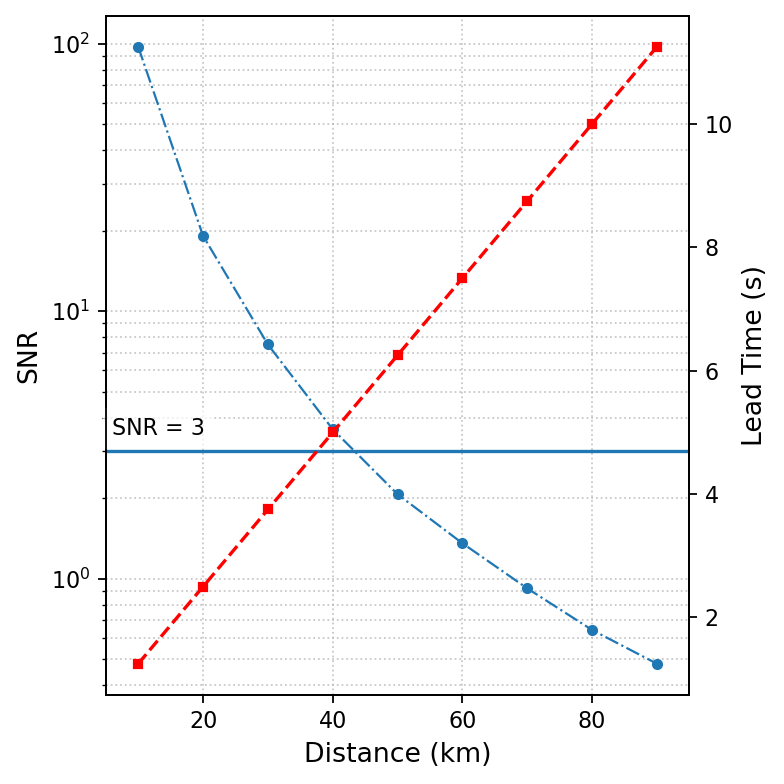}
    \caption{Lead time versus distance.}
    \label{fig:EQ_b}
\end{subfigure}
\caption{(a) Predicted sensitivity of CHRONOS overlaid with prompt gravity signals generated by a Mw 5.2 earthquake at various detector distances. (b) Signal-to-Noise ratios (blue circles) and lead-time (red squares) at various detector distances. SNR above SNR=3(horizontal line) may be considered detectable.}
\label{fig:EQ}
\end{figure}

Figure~\ref{fig:EQ_a} indicates that at distances less than $90$km, the expected signal amplitude falls within the CHRONOS sensitivity band in the sub-Hz range. To show detectability, Figure~\ref{fig:EQ_b} contains the signal-to-noise ratio(SNR)(blue circles) and lead time(red squares) at various distances, assuming an average P-wave speed of $8.0$ km/s, showing that lead time increases while SNR decreases with distance. At $40$ km, SNR is $3.62$ with a lead time of $5.00$ seconds at P-wave speed of $8.0$ km/s. For P-wave speed ranging from $5.8$ to $13.7$ km/s \cite{bormann2012seismic}, still at $40$ km, we calculated detections that were faster compared to conventional seismometers by 2.92 to 6.90 seconds. Detecting these signals offers a possible route for early earthquake warning and enables direct observation of time-varying gravitational fields caused by terrestrial mass redistribution.

\newpage
\section{Conclusions}
Our study indicates that CHRONOS has the potential to fill a gap in the largely unexplored 0.1–10 Hz gravitational wave detection band, achieving sensitivity $h \sim 10^{-18}\,\mathrm{Hz^{-1/2}}$ around $2\,\mathrm{Hz}$, competitive with existing detectors. By addressing quantum, thermal, and environmental noise limitations, CHRONOS not only advances ground-based interferometry capabilities, but also opens new opportunities for geophysical applications, such as earthquake early-warning. The current model predicts that earthquakes with magnitude $5.2$ within $40$ km may be detected with SNR greater than $3$ with a lead time of $5$ seconds. These observations of earthquake-induced prompt gravitational signals not only demonstrate a novel application of interferometric gravity sensing but also aid validation of environmental noise models for future detectors, providing strong motivation for continued research and development of sub-Hz interferometric detectors. 

\begin{acknowledgments}
We would like to express our sincere gratitude to Y-C.Lin, A.Markoshan, M.Hasegawa, T.Kanayama, and M.Hazumi for their valuable discussions and continuous support throughout this work. We also acknowledge the support and collaborative environment provided by the Department of Physics and the Center for High Energy and High Field (CHiP) at National Central University, the Institute of Physics, Academia Sinica, the National Institute of Physics, University of the Philippines Diliman, as well as Taiwan Semiconductor Research Institute (TSRI). Y.I. is supported by the National Science and Technology Council (NSTC) of Taiwan under Grant No. 114-2112-M-008-006, and by Academia Sinica under Grant No. AS-TP-112-M01.
\end{acknowledgments}

\bibliographystyle{apsrev4-2}
\bibliography{bibfile}

\end{document}